\documentclass[twocolumn,conference]{IEEEtran}
\usepackage[T1]{fontenc}
\usepackage[latin9]{inputenc}
\usepackage{units}
\usepackage{amsmath}
\usepackage{amsthm}
\usepackage{amssymb}
\usepackage{graphicx}
\usepackage[unicode=true,
 bookmarks=true,bookmarksnumbered=true,bookmarksopen=true,bookmarksopenlevel=1,
 breaklinks=false,pdfborder={0 0 0},pdfborderstyle={},backref=false,colorlinks=false]
 {hyperref}
\hypersetup{pdftitle={Your Title},
 pdfauthor={Your Name},
 pdfpagelayout=OneColumn, pdfnewwindow=true, pdfstartview=XYZ, plainpages=false}

\makeatletter

\providecommand{\tabularnewline}{\\}

\theoremstyle{plain}
\newtheorem{thm}{\protect\theoremname}

\usepackage[caption=false,font=footnotesize]{subfig}

\makeatother

\providecommand{\theoremname}{Theorem}

\begin{document}
\title{Unnormalized Measures in Information Theory}
\author{\IEEEauthorblockN{Peter Harremoës}\IEEEauthorblockA{Niels Brock\\
Copenhagen Business College\\
Copenhagen\\
Denmark\\
Email: harremoes@ieee.org}}
\maketitle
\begin{abstract}
Information theory is built on probability measures and by definition
a probability measure has total mass 1. Probability measures are used
to model uncertainty, and one may ask how important it is that the
total mass is one. We claim that the main reason to normalize measures
is that probability measures are related to codes via Kraft's inequality.
Using a minimum description length approach to statistics we will
demonstrate with that measures that are not normalized require a new
interpretation that we will call the Poisson interpretation. With
the Poisson interpretation many problems can be simplified. The focus
will shift from from probabilities to mean values. We give examples
of improvements of test procedures, improved inequalities, simplified
algorithms, new projection results, and improvements in our description
of quantum systems. 
\end{abstract}

\section{Introduction}

In 1933 Kolmogorov gave a firm foundation to probability theory by
defining a \emph{probability measure} as a measure for which the total
mass is 1 \cite{Kolmogorov1933}. After his article, it was possible
to study probability theory as a purely mathematical topic and that
has given rise to a tremendious number of results with both practical
applications and applications in other brances of mathematics. 

In his 1933 article it is stated as an axiom that the total probability
mass is one. One may look for theorems that justify that the total
mass should be one. a number of important theorems are formulated
in terms of probability measures, but a closer inspection reveals
that they are essentially about measures with finite total mass that
may be normalized to be one. For instance any compact group has a
unique invariant probability measure \cite{Haar1933}, but the more
general theorem is that any locally compact group has a left invariant
measure (Haar measure) that is unique except for a multiplicative
constant. Since a compact group has finite Haar measure one can normalize
the measure, but the theorem does not tell why one would prefer to
normalize the measure so that the total mass is one rather than say
two or three.

One of the very few theorems that really gives a preference for measures
of total mass one, is Kraft's Inequality.
\begin{thm}[{Modified version of \cite[Thm. 1]{Harremoes2013}}]
Let $\ell:\mathbb{A}\to\mathbb{R}$ be a function. Then the function
$\ell$ satisfies Kraft's Inequality 
\[
\sum_{a\in\mathbb{A}}\mathrm{e}^{-\ell\left(a\right)}\leq1
\]
if and only if for any $\epsilon>0$ there exists an integer $n$
and a uniquely decodable fixed-to-variable length block code $\kappa:\mathbb{A}^{n}\to\mathbb{B}^{*}$
such that 
\[
\left|\bar{\ell}\left(a^{n}\right)-\frac{1}{n}\sum_{i=1}^{n}\frac{\ell\left(a_{i}\right)}{\ln\left(\beta\right)}\right|\leq\epsilon
\]
 where $\bar{\ell}_{\kappa}\left(a^{n}\right)$ denotes the length
$\ell_{\kappa}\left(a^{n}\right)$ divided by $n$. The uniquely decodable
block code can be chosen to be prefix free.
\end{thm}
If we define $\mu\left(a\right)=\mathrm{e}^{-\ell\left(a\right)}$
we get a one-to-one correspondence be abstract codelength functions
$\ell$ and sub-probability measures $\mu.$ 
\begin{thm}
Let $\mu$ denote a measure on the discrete alphabet $\mathbb{A}$
and let $A\subseteq\mathbb{A}$ and assume that $\mu\left(A\right)<\infty.$
Then the abstract code length function $\ell$ that minimize the mean
code length 
\[
\sum_{a\in A}\mu\left(a\right)\ell\left(a\right)
\]
 is 
\[
\ell\left(a\right)=-\ln\left(\mu\left(a\mid A\right)\right)
\]
 where 
\[
\mu\left(a\mid A\right)=\frac{\mu\left(a\right)}{\sum_{\alpha\in A}\mu\left(\alpha\right)}.
\]
\end{thm}
We see that probability measures are naturally associated with the
problem of optimizing code length. In particular this theorem justify
the use of conditional probaiblities even in cases where the measure
$\mu$ is not finite. 

Since information theory via Kraft's inequality may justify that the
total mass should be one, we may ask if unnormalized measures are
of any use in information theory. In this note we will review to what
extend it is really needed to assume that our measures are normalized
so that the total mass is 1. New results will be developed, but most
of the proofs are omitted. Some of the proofs are simple modifications
of previously published proof.

\section{The Poisson interpreation}

If $P$ and $Q$ are probability measures and if $P$ is absolutely
continuous with respect to $Q$ information divergence is defined
as
\begin{align}
D\left(P\Vert Q\right) & =\sum_{i}p_{i}\ln\left(\frac{p_{i}}{q_{i}}\right)\label{eq:KLformula}\\
 & =\sum_{i}p_{i}\ln\left(\frac{1}{q_{i}}\right)-\sum_{i}p_{i}\ln\left(\frac{1}{p_{i}}\right)\nonumber 
\end{align}
which means that information divergence is the difference between
mean value of the optimal code length $\ln\left(\frac{1}{q_{i}}\right)$
corresponding to $P$ and the optimal code length corresponding to
$Q$ where both means are calculated with respect to $P.$ Next we
will extend divergence from probability vectors to arbitrary positive
vectors.

We define a \emph{Bernoulli random vector} as a random vector, which
equals one of the base vectors with probability 1. If $\vec{X}$ is
a Bernoulli random vector then $E\left(\vec{X}\right)$ is a probability
vector. We say that $\vec{Y}$ is a \emph{Bernoulli sum} if $\vec{Y}$
is a sum of independent Bernoulli random vectors. Let $Po\left(\lambda\right)$
denote the Poisson distribution with mean value $\lambda.$ If $\vec{\lambda}=\left(\lambda_{1},\lambda_{2},\dots,\lambda_{k}\right)$
then the multivariate Poisson distribution $Po\left(\vec{\lambda}\right)$
is defined as the product measure $\bigotimes_{i=1}^{k}Po\left(\lambda_{i}\right).$
With this terminology we can generalize a result from \cite{Harremoes2001c}:
\begin{thm}
The maximum entropy distribution of Bernoulli sums $\vec{Z}$ satisfying
$E\left(\vec{Z}\right)=\vec{\lambda}$ is the multivariate Poisson
distribution with mean value $\vec{\lambda}.$
\end{thm}
The idea of \emph{thinning} is that a proportion of the observations
are discarted. In an $\alpha$-thinning an observation is kept with
probability $\alpha$. We note that the $\alpha$-thinning of the
binomial distribution $bin\left(n,p\right)$is the binomial distribution
$bin\left(n,\alpha\cdot p\right)$ and the $\alpha$-thinning of the
Poisson distribution $Po\left(\lambda\right)$ is the Poisson distribution
$Po\left(\alpha\cdot\lambda\right).$ We will estend the notion of
thinning of random variables to \emph{thinning of random vectors}.
If $\text{\ensuremath{\vec{Y}=\vec{X}_{1}+\vec{X}_{2}+\dots+\vec{X}_{n}}}$then
the $\alpha$-thinning of the distribution of $\vec{Y}$ is the distribution
of $B_{1}\cdot\vec{X}_{1}+B_{2}\cdot\vec{X}_{2}+\dots+B_{n}\cdot\vec{X}_{n}$
where $B_{1},B_{2},\dots,B_{n}$ are iid Bernoulli random variables
with succes probability $\alpha.$ If the $P$ is the distribution
of $\vec{Y}$ then the distribution of the $\alpha$-thinning of $\vec{Y}$
is denoted $T_{\alpha}\left(P\right)$. This vector thinning essentially
thins each of the coordinates of the vector independently. These definitions
allow us to prove vecter versions of results from \cite{Harremoes2008c,Harremoes2010c}.
\begin{thm}[Law of thin vectors]
Let $\vec{Z}$ be a random vectors with values in $\mathbb{N}_{0}^{k}$
and with $E\left(\vec{Z}_{i}\right)=\vec{\lambda}$. If $\vec{Z}$
has distribution $P$, then $T_{\nicefrac{1}{n}}\left(P^{*n}\right)$
has mean value $\vec{\lambda}.$ Further $T_{\nicefrac{1}{n}}\left(P^{*n}\right)$
convergences to the maximum entropy distribution in total variation.
If $\vec{Z}$ is a Bernoulli sums then $D\left(\left.T_{\nicefrac{1}{n}}\left(P^{*n}\right)\right\Vert Po\left(\vec{\lambda}\right)\right)\to0$
and $H\left(T_{\nicefrac{1}{n}}\left(P^{*n}\right)\right)\to H\left(Po\left(\vec{\lambda}\right)\right)$
for $n\to\infty.$ 
\end{thm}
\begin{thm}
If the probability vectors $P=\vec{\lambda}$ and $Q=\vec{\mu}$ are
distributions of Bernoulli random vectors, then 
\begin{align*}
D\left(P\Vert Q\right) & =D\left(\left.T_{\nicefrac{1}{n}}\left(P^{*n}\right)\right\Vert T_{\nicefrac{1}{n}}\left(Q^{*n}\right)\right)\\
 & =D\left(\left.Po\left(\vec{\lambda}\right)\right\Vert Po\left(\vec{\mu}\right)\right).
\end{align*}
\end{thm}
We have 
\begin{align*}
D\left(\left.Po\left(\vec{\lambda}\right)\right\Vert Po\left(\vec{\mu}\right)\right) & =\sum_{i=1}^{k}D\left(\left.Po\left(\lambda_{i}\right)\right\Vert Po\left(\mu_{i}\right)\right)\\
 & =\sum_{i=1}^{k}\lambda_{i}\ln\left(\frac{\lambda_{i}}{\mu_{i}}\right)-\left(\lambda_{i}-\mu_{i}\right).
\end{align*}
For probability vectors the original formula for divergence is recovered,
but the interpretation is different. Information divergence is the
difference in mean code length between the code corresponding to $\bigotimes_{i}Po\left(\lambda_{i}\right)$
and $\bigotimes_{i}Po\left(\mu_{i}\right)$ where the mean is calculated
with respect to $\bigotimes_{i}Po\left(\lambda_{i}\right).$ With
this interpretation we can easily extend the definition of $D\left(P\Vert Q\right)$
to situations where $P$ and $Q$ are general measures rather than
probability measures. We will write $D\left(\left.\vec{\lambda}\right\Vert \vec{\mu}\right)=\sum_{i=1}^{k}\lambda_{i}\ln\left(\frac{\lambda_{i}}{\mu_{i}}\right)-\left(\lambda_{i}-\mu_{i}\right)$
and note that in this formula $\vec{\lambda}$ and $\vec{\mu}$ may
be arbitrary vectors with positve entries, i.e. $\vec{\lambda}$ and
$\vec{\mu}$ are arbitrary measures. 

On the set of measures we have two basic operations. The measures
$P$ and $Q$ can be \emph{added}. The interpretation is that two
experiments are performed independently. For each category $i$ we
get counts $X_{i}$ and $Y_{i}$, and the results are combined by
adding the counts to $X_{i}+Y_{i}.$ If $\alpha\in\left[0,1\right]$
then a measure $P$ can be \emph{multiplied} by $\alpha$. The interpretation
is that if one has obtained a count $X_{i}$ in category $i$ then
$X_{i}$ is replaced by $Z_{i}\sim bin\left(X_{i},\alpha\right)$corresponding
to removing each observations with probability $\alpha.$ In information
theory this corresponds to concatenation and applying a deletion channel.

The advantage of using Poisson distributions is that this class of
distributions is closed under repetition and thinning. The interpretation
of the measure $P$ is that it is a vector $\left(p_{1},p_{2},\dots p_{k}\right)$
which the mean value of a rando vector of counts with distribution
$\bigotimes_{i}Po\left(p_{i}\right).$

\section{Testing Goodness-of-Fit}

Here we will look at the consequences for testing Good-of-Fit in one
of the simplest possible setups. We will test if coin is fair, and
we perform an experiment where we count the number of heads $X$ and
the number of tails $Y$ after tossing the coin a number of times.
Our nul-hypothesis is that there is symmetry between heads and tails.
Here we will compare the analysis for the case when we have observed
$X=\ell$ and $Y=m$.

\subsection{Classical analysis}

Classically one will fix the number of tosses so that $X+Y=n$, and
assume that $X$ has a binomial distribution with success probability
$p.$ The classical nul-hypothesis is that $p=\nicefrac{1}{2}$.

\begin{figure}[tbh]
\begin{centering}
\includegraphics[viewport=20bp 45bp 289bp 220bp,clip,width=5cm,height=5cm]{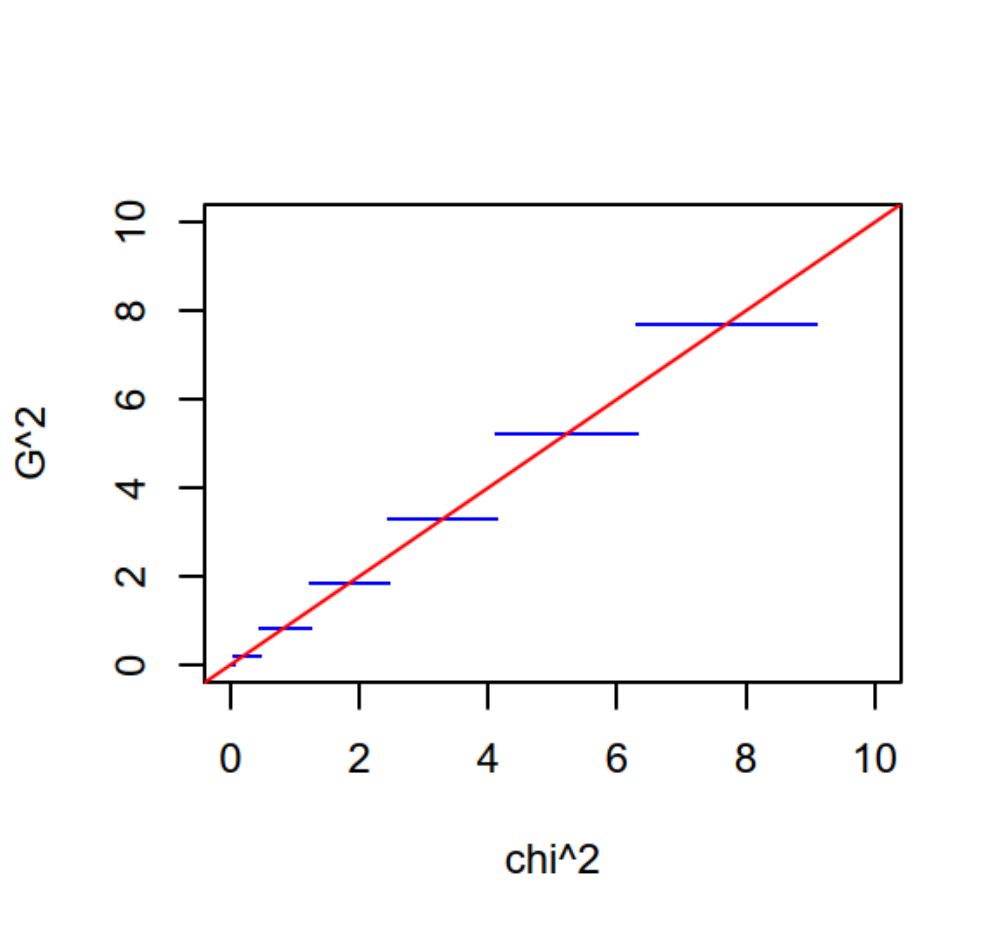}
\par\end{centering}
\caption{\label{fig:QQbinom}QQ-plot of the $\chi^{2}$-distribution with $df=1$
against the distribution of the $G^{2}$-statistic for testing $p=\nicefrac{1}{2}$
in $bin\left(20,p\right)$.}
\end{figure}

The maximum likelihood estimate of $p$ is $\nicefrac{\ell}{n}.$
The divergence is 
\begin{multline*}
D\left(\left.bin\left(n,\nicefrac{\ell}{n}\right)\right\Vert bin\left(n,\nicefrac{1}{2}\right)\right)\\
=n\cdot\left(\frac{\ell}{n}\ln\left(\frac{\nicefrac{\ell}{n}}{\nicefrac{1}{2}}\right)+\frac{m}{n}\ln\left(\frac{\nicefrac{m}{n}}{\nicefrac{1}{2}}\right)\right).
\end{multline*}
We introduce the signed log-likelihood as
\begin{multline*}
G_{n}\left(x\right)=\\
\begin{cases}
-\left(2\cdot D\left(\left.bin\left(n,\nicefrac{x}{n}\right)\right\Vert bin\left(n,\nicefrac{1}{2}\right)\right)\right)^{\nicefrac{1}{2}}, & \textrm{if }x<\nicefrac{n}{2};\\
+\left(2\cdot D\left(\left.bin\left(n,\nicefrac{x}{n}\right)\right\Vert bin\left(n,\nicefrac{1}{2}\right)\right)\right)^{\nicefrac{1}{2}}, & \textrm{if }x\geq\nicefrac{n}{2}.
\end{cases}
\end{multline*}
In \cite[Cor 7.2]{Harremoes2016} it is proved that
\[
\Pr\left(X<k\right)\leq\Phi\left(G_{n}\left(k\right)\right)\leq\Pr\left(X\leq k\right).
\]
A QQ plot with a Gaussian distribution on the first axis and the distribution
of $G\left(X\right)$ on the second axis one gets a stair with horisontal
steps each intersecting the line $x=y$ corresponding to a perfect
match between the distribution of $G\left(X\right)$ and a standard
Gaussian distribution. If we square $G\left(X\right)$ we get 2 times
divergence, which is often called the $G^{2}$-statistic. Due to symmetry
between head and tail the intersection property is also satisfied
when the distribution of the $G^{2}$-statistic is compared with a
$\chi^{2}$-distribution \cite{Harremoes2012}. This is illustrated
in Figure \ref{fig:QQbinom}. Instead of using the Gaussian approximation
one could calculate tail probabilities exactly (Fisher's exact test),
but as we shall see below we can do better.

\subsection{Poisson analysis\label{subsec:Poisson-analysis}}

In our setup we assume that $X$ and $Y$ are independent Poisson
distributed random variables with mean values $\lambda$ and $\mu$
respectively. In our analysis the nul-hypothesis is that $\lambda=\mu.$

\begin{figure}[tbh]
\begin{centering}
\includegraphics[viewport=30bp 45bp 289bp 220bp,clip,width=5cm,height=5cm]{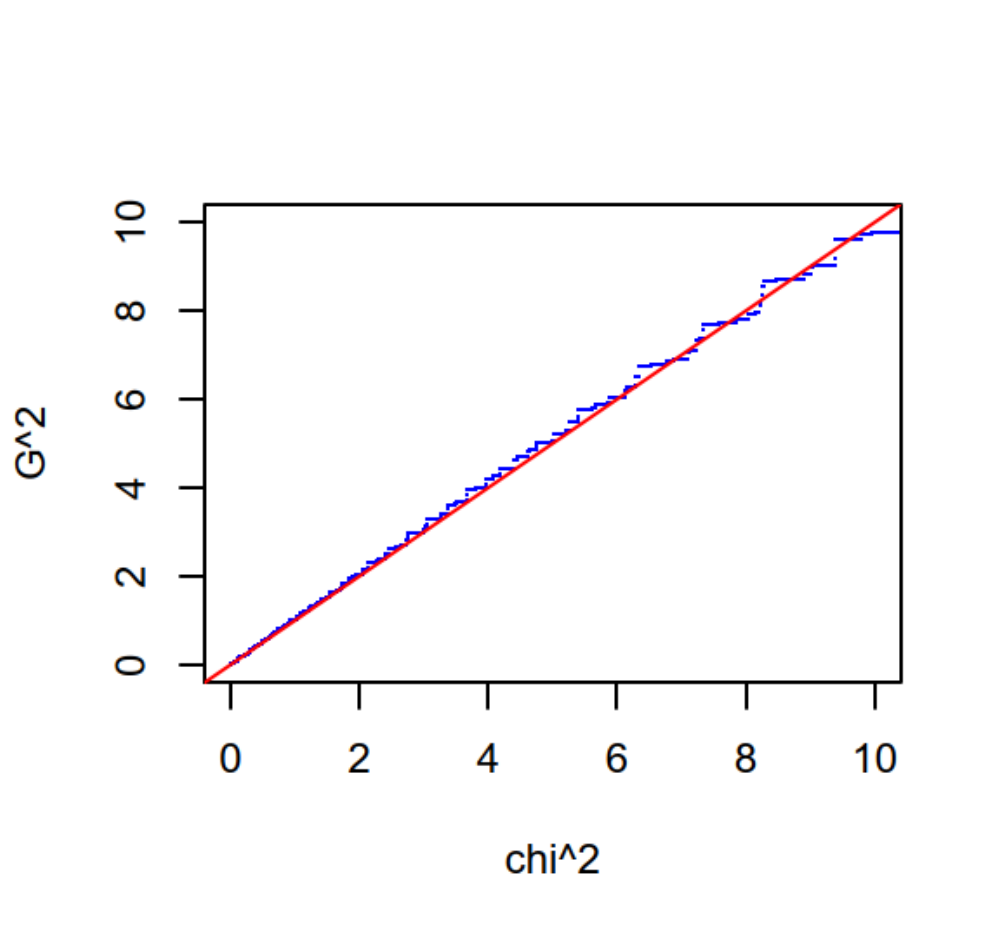}
\par\end{centering}
\caption{\label{fig:QQ-plot-Poisson}QQ-plot of the $\chi^{2}$-distribution
with $df=1$ against the distribution of the $G^{2}$-statistic for
testing $\lambda=\mu$ based on $\hat{\lambda}+\hat{\mu}=20.$}
\end{figure}

Since $X+Y\sim Po\left(\lambda+\mu\right)$ the maximum likelihood
estimate of $\lambda+\mu$ is $\ell+m.$ Hence the estimate of $\lambda$
and $\mu$ are $\left(\ell+m\right)/2.$ Here we define $N=X+Y$ and
$n=\ell+m.$ We calculate the divergence 
\begin{multline*}
D\left(Po\left(\ell\right)\otimes Po\left(m\right)\Vert Po\left(\frac{n}{2}\right)\otimes Po\left(\frac{n}{2}\right)\right)\\
=n\left(\frac{\ell}{n}\ln\left(\frac{\nicefrac{\ell}{n}}{\nicefrac{1}{2}}\right)+\frac{m}{n}\ln\left(\frac{\nicefrac{m}{n}}{\nicefrac{1}{2}}\right)\right).
\end{multline*}
i.e. the same expression as in the classical analysis. Since $X$
is binomial given that $X+Y=n$ we have 
\begin{multline*}
\Pr\left(X<k\mid N=n\right)\leq\Phi\left(G_{n}\left(k\right)\right)\\
\leq\Pr\left(X\leq k\mid N=n\right),
\end{multline*}
Since the distribution of $\left(G_{N}\left(X\right)\right)^{2}$
is close to a $\chi^{2}$-distribution under the condition $N=n$
the same is true for $\left(G_{N}\left(X\right)\right)^{2}$ when
we take the mean value over $N.$ Since each of the steps intersect
the streight line near the mid point of the step the effect of taking
the mean value with respect to $N$ is that the steps to a large extend
cancel out as illustrated in Figure \ref{fig:QQ-plot-Poisson}. If
one were testing a nul-hypothesis with less symmetry one will essentially
get the same result except that the left tail and the right tail of
the signed log-likelihood should be handled separately. 

For statistical analysis one should not fix the sample size before
sampling. A better procedure is to sample for a specific time so that
the sample size becomes a random variable. Often in practice this
is how sampling takes place and if the sample size is really random
it may even be misleading to analyze data as in the classical setup.

\section{Quantum states}

We have argued that it is often useful to consider vectors of mean
values rather than probability vectors. Since the notion of quantum
states and measurements are defined in terms of probability vectors
\cite{Holevo1982} it is relevant to define these concepts in terms
of mean values rather than probabilities. The limited space available
in this paper does not allow us to go into a proper treatment of this
problem. Instead we vill just give an example of how the shift in
focus get us closer to the experimental reality than using the model
based on probabilities.

\begin{figure}[tbh]
\begin{centering}
\includegraphics[scale=0.3]{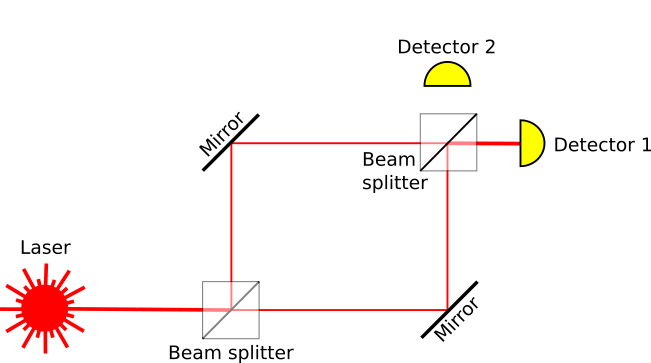}
\par\end{centering}
\caption{\label{fig:Mach-Zehnder-Interferometer}Mach-Zehnder Interferometer.
Here both paths are open and Detector 2 does not detect any photons
due to interference.}

\end{figure}

Often the double slit experiment is presented to illustrate how quantum
mechanics differ from classical mechanics. A simplified version of
the double slit experiment is the Mach-Zehnder interferometer where
photons are emitted by a laser. If the photons can take both paths
then interference implies that only Detector 1 will detect photons
as illustrated in Figure \ref{fig:Mach-Zehnder-Interferometer}. If
one of the paths is blocked then no interference takes place and photons
will be detected at both detector 1 and detector 2. In most descriptions
of the double slit experiment it is explained that the interference
takes place even if ``the intensity of the laser is so low that it
only emits single photons''. Typically it is stated that the probability
for detection at Detector 1 is 1 if both paths are possible, but if
one of the paths is blocked then detection of the photon at each of
the detectors 1 or 2 have probability $\nicefrac{1}{2}$. In an attempt
to model quantum mechanics by concepts related to probability theory
we get a description that appear paradoxical.

A real laser emits light that can be described by a coherent state
and in this case the photon detections follows a Poisson process.
This has been used to build cryptographic protocols \cite{Hasan2018}.
A more realistic description of the Mach-Zehnder interferometer is
as follows. If both paths are open then detector 1 detects photons
according to a Poisson process and detector 2 does not detect any
photons. If one of the paths is blocked then dector 1 detects photons
with intensity $\lambda$ per second and detector 2 detects photons
with intensity $\mu$ per second. For a perfect bean splitter symmetry
we have $\lambda=\mu.$ One can make the experiment and test the hypothesis
$\lambda=\mu$ and in Subsection \ref{subsec:Poisson-analysis} we
discussed how this can be done most efficiently. The result is that
the hypothesis $\lambda=\mu$ can be accepted even if the intensity
of the laser is low. 

The idea of single photons is only imposed in hindsight after the
detection has taken place. In order to get a probability measure one
condition on the event that a specific total number of photons have
been observed. It should be noted that the conditioning is part of
the measurement or of the processing of the data after the measurement.
It is not part of the preparation of the system. With a correct description
in terms of intensties of Poisson processes the paradoxical description
involving paths of individual photons dissolves.

\section{Projections based on f-divergences}

Conditioning is a special case an information projection\cite{Harremoes2002c},
and maximum likelihood estimation is a special case of a reversed
information projection. Both information divergence and reverse information
divergence are examples of $f$-divergences, and for this reason our
next results are formulated for $f$-divergences.

Let $f:\left]0;\infty\right[\to\mathbb{R}$ denote a convex function
such that $f\left(x\right)\geq0$ with equality if $x=1$. Define
\begin{align*}
f\left(0\right) & =\lim_{x\to0}f\left(x\right),\\
f'\left(\infty\right) & =\lim_{x\to\infty}\frac{f\left(x\right)}{x}.
\end{align*}
We introduce the convention that
\[
f\left(\frac{x}{0}\right)\cdot0=f'\left(\infty\right)\cdot x.
\]
For finite measures $P$ and $Q$ the $f$-divergence is defined as
a number in $\left[0,\infty\right]$ given by
\[
D_{f}\left(P,Q\right)=\sum_{i}f\left(\frac{p_{i}}{q_{i}}\right)\cdot q_{i}.
\]

For information divergence $f\left(x\right)=x\cdot\ln\left(x\right)-\left(x-1\right)$
with $f'\left(\infty\right)=\infty.$ For reversed information divergence
$f\left(x\right)=-\ln\left(x\right)+x-1$ with $f'\left(\infty\right)=1.$

Let $C$ denote a convex set and let $Q$ denote a distribution. Then
define 
\[
D\left(C,Q\right)=\inf_{p\in C}D_{f}\left(P,Q\right).
\]
A sequence $P_{n}\in C$ is said to be asymptotically optimal if 
\[
D_{f}\left(P_{n},Q\right)\to D_{f}\left(C,Q\right)
\]
 for $n\to\infty.$
\begin{thm}
Assume that $f$ is strictly convex and assume that $C\subseteq M_{+}$
and $Q\in M_{+}^{1}.$ If $D_{f}\left(C,Q\right)<\infty$ then there
exists a bounded measure $Q^{*}\in M_{+}$ such that for an asymptotically
optimal sequence $P_{n}$ we have $P_{n}\left(i\right)\to Q^{*}\left(i\right)$
for $Q$-almost any $i.$ In particular, the measure $Q^{*}$ is unique
$Q$-almost surely.
\end{thm}
\begin{thm}
Assume that $C\subseteq M_{+}^{1}$ and $Q\in M_{+}^{1}$ and $D_{f}\left(C,Q\right)<\infty$.
If $f$ is strictly convex and $f'\left(\infty\right)=\infty$ then
the projection $Q^{*}$ is a unique probability measure.
\end{thm}
\begin{thm}
Let $Q$ denote a probability measure and let $g:\mathbb{N}\to\mathbb{R}$
denote a positive function and assume that mean value $\sum g\left(i\right)\cdot q_{i}$
is $\mu.$ Assume that $0<\tilde{\mu}<\mu$. Let $C$ denote the compact
set of measures $P$ for which $\sum g\left(i\right)\cdot p_{i}\leq\tilde{\mu},$
and assume that $D_{f}\left(C,Q\right)<\infty.$ If $f'\left(\infty\right)<\infty$
then the projection of $Q$ on $C$ satisfies 
\begin{align*}
\sum g\left(i\right)\cdot q_{i}^{*} & =\tilde{\mu},\\
\sum q_{i}^{*} & <\sum q_{i}.
\end{align*}
In addition $Q^{*}$ is absolutely continuous with respect to $Q.$
\end{thm}
The last result demonstrates that unnormalized measures naturally
appear as reversed information projections on sets of probability
measures. This is of particular relevance for new methods for statistical
testing where the tests are based on $E$-values rather than the usual
$P$-values \cite{Gruenwald2021}.

\section{Some improved projection inequalities}
\begin{thm}
Let $Q$ denote a measure and let $X$ denote a random variable with
$E_{Q}\left(f\left(X\right)\right)=0.$ If $E_{Q}\left(f\left(X\right)^{2}\right)=1$
and $E_{Q}\left(\left(f\left(X\right)\right)^{3}\right)>0.$ If there
exist $\beta<0$ such that $Z\left(\beta\right)=\int\exp\left(\beta\cdot x\right)\,\mathrm{d}Qx<\infty,$then
there exists $\epsilon>0$ such that for all measures $P$ with $E_{p}\left(X\right)\in\left[-\epsilon,0\right]$
we have the inequality 
\[
D\left(P\Vert Q\right)\geq\frac{1}{2}\left(E\left(X\right)\right)^{2}.
\]
\end{thm}
The conditions can be applied for the Gaussian distribution and associated
normalized Hermite polynomials. It also holds for a Poisson distribution
and associated Poisson-Charlier polynomials. It also holds for Binomial
distributions and associated Kravshuk polynomials. This can be used
to give bounds on rate on convergence in the Central Limit Theorem
\cite{Harremoes2003a,Harremoes2005c,Harremoes2006b}, in the Law of
Thin Numbers \cite{Harremoes2004,Harremoes2006c,Harremoes2007b,Harremoes2008c,Harremoes2010c,Harremoes2016b},
and in approximations of hyper geometric distributions by binomial
distributions \cite{Harremoes2020a}. The theorem improves previous
results by not assuming that $P$ is a probability measure. For Gaussian
distributions, Poisson distributions, and binomial distributions one
may even drop the condition $E_{p}\left(X\right)\in\left[-\epsilon,0\right]$
as long as $E_{p}\left(X\right)\leq0$ and the the orthogonal polynomials
have sufficiently small order, but each case require special techniques
and the proofs are computationally involved, but actually it gies
some slight simplifications if we drop the condition that the measures
should be normalized.. 

\section{Alternating minimization}

Several algorithms in information theory involve information projections,
and some of them can be simplified by dropping the condition that
we should stick to probability measures. One example is alternating
minimization.

Let $Q$ denote a probability measure and let $C_{1},C_{2},\dots,C_{k}$
denote a sequence of convex sets of probability measures and let $C$
denote their intersection. In order to find the information projection
of $Q$ on $C$ one can find the projection $Q_{1}$ of $Q$ on $C_{1}.$
Then one projects $Q_{1}$ on $C_{2}$ leading a projection $Q_{2}$
and so forth taking the sets $C_{1},C_{2},\dots,C_{k}$ in cyclic
order. Then the sequence of projections $Q_{1},Q_{2},\dots$ will
converge to the projection of $Q$ on $C$. Different versions of
this algorithm have many important applications. Therefore it is useful
if we can simplify this algorithm and speed up the rate of conversion.

Assume that each of the sets $C_{i}$ are given by a mean value constraint
of the form 
\[
C_{i}=\left\{ P\mid\sum_{j}f_{i}\left(j\right)\cdot p_{j}=\mu_{i}\wedge\sum_{j}p_{j}=1\right\} .
\]
 Let 
\[
\tilde{C}_{i}=\left\{ P\mid\sum_{j}f_{i}\left(j\right)\cdot p_{j}=\mu_{i}\right\} .
\]
 Let $\tilde{C}_{0}=\left\{ P\mid\sum_{j}p_{j}=1\right\} $ and note
that $C_{i}=\tilde{C}_{i}\cap\tilde{C}_{0}.$ Instead of projecting
on $C_{i},I=1,2,\dots,k$ in cyclic order one can project on $\tilde{C}_{i},i=0,1,\dots,k$
in cyclic order. This simplifies each of the projections.

In order to accelerate the algorithm we may replace the function $f_{0},f_{1},f_{2},\dots,f_{k}$
by functions that are orthogonal with respect to $Q.$ Such orthogonal
functions are easily calculated using the Gram-Smith procedure.

\section{Discussion}

As we have seen one may replace probability measures by more general
measures without loosing the interpreation of information divergence
as the mean difference in code length. It may sometimes require that
a probability $p_{i}$ in a probability vector $P=\left(p_{1},p_{2},\dots,p_{k}\right)$
are interpreted as a mean value of a count that is Poisson distributioned.
That means that instead of having a count that may assume the two
values 0 and 1 we get random variables that may assume any values
in $\mathbb{N}_{0}$. This is definitely more abstract, which may
seem complicate the foundation of probability theory, but often it
allow us more freedom if we can more freely swich between considering
the probability vector $\left(p_{1},p_{2},\dots,p_{k}\right)$ as
describing independent events and as describing mutually excluding
events. The translation forth and back is to consider a vector of
counts $\left(Z_{1}Z_{2},\dots Z_{k}\right)$. If they are mutually
exclusive and Bernoulli then according to the Law of Thin Numbers
the thinned sum of independent copies of this vector is approximately
Poisson distributed and if $\left(Z_{1}Z_{2},\dots Z_{k}\right)$
is Poisson distributed then conditioning on $Z_{1}+Z_{2}+\dots+Z_{k}=1$
leads to a multinomial distribution.

\medskip{}

\begin{tabular}{|c|c|}
\hline 
\textbf{Standard interpretation} & \textbf{Poisson interpretation}\tabularnewline
\hline 
Probability measure & Measure\tabularnewline
\hline 
Multinomial distribution & Poisson distribution\tabularnewline
\hline 
Probability & Mean value\tabularnewline
\hline 
KL-divergence & $f$-divergence\tabularnewline
\hline 
Code & Conditional probability\tabularnewline
\hline 
Product measure & Sum of measures\tabularnewline
\hline 
\end{tabular}

By switching the focus from probabilities to mean values and integrals
we also get in accordance with the following formulation of the Dutch
Book Theorem.
\begin{thm}[\cite{Harremoes1993,Harremoes2009c}]
Let $X_{1},X_{2},\dots,X_{n}$ denote random variables (i.e. functions)
on a finite sample space $\Omega.$ Then either there exists positve
weights $s_{1},s_{2},\dots,s_{n}$ such that the linear combination
\[
X=s_{1}X_{1}+s_{2}X_{2}+\dots+s_{n}X_{n}
\]
satisfies $X\left(\omega\right)<0$ for all $\omega\in\Omega$ or
there exists a measure $\mu$ on $\Omega$ such that 
\[
\int X_{i}\left(\omega\right)\,\mathrm{d}\mu\omega\geq0
\]
 for all $i=1,2,\dots,n$. 
\end{thm}
A complete theory about how the ideas presented in this paper can
be used not only to extend classical probability theory, but may replace
classical probability theory as a foundation of our understanding
of randomness, uncertainty, and quantum information theory, is work
in progress.

\newpage{}

\bibliographystyle{IEEEtran}
\bibliography{C:/Users/peth.NB/Documents/BibTeX/database1}

\end{document}